\documentclass[prl,aps,showpacs,twocolumn,longbibliography]{revtex4}

\usepackage{amssymb}
\usepackage{amsmath}
\usepackage{mathtools}
\usepackage[dvipdfm]{hyperref}
\usepackage{hyperref}
\usepackage{stmaryrd}
\usepackage{float}

\allowdisplaybreaks
\begin{document}

\title{Quantifying the phase of quantum states}
\author{Jianwei Xu}
\email{xxujianwei@nwafu.edu.cn}
\affiliation{College of Science, Northwest A$\&$F University, Yangling, Shaanxi 712100,
China}

\begin{abstract}
Phase is a basic ingredient for quantum states since quantum mechanics uses
complex numbers to describe quantum states. In this letter, we introduce a
rigorous framework to quantify the phase of quantum states. To do so, we
regard phase as a quantum resource, and specify the free states and free
operations. We determine the conditions a phase measure should satisfy and
provide some phase measures. We also propose the notion of intrinsic phase
for quantum states.
\end{abstract}

\maketitle

\emph{Introduction.}---Phase is a basic concept in quantum states since quantum mechanics uses
complex numbers to describe quantum systems. One of the postulates of
quantum mechanics is that a quantum system is described by a complex Hilbert
space. We assume the Hilbert space is $C^{d}$ with $d$ the dimension and $C$
the set of complex numbers. A pure state is represented by a normalized
vector $|\psi \rangle $ in $C^{d}.$ Fix an orthonormal basis $\{|j\rangle
\}_{j=1}^{d}$ of $C^{d},$ we can explicitly write out $|\psi \rangle $ as $%
|\psi \rangle =\sum_{j=1}^{d}c_{j}|j\rangle $ with $c_{j}=\langle j|\psi
\rangle $ and $\sum_{j=1}^{d}|c_{j}|^{2}=1.$ Write each $c_{j}$ in the polar
form $c_{j}=|c_{j}|e^{i\alpha _{j}}$ with $i=\sqrt{-1}$, $|c_{j}|$ the
modulus of $c_{j}$, $\alpha _{j}\in R$, $R$ the set of real numbers, then
there appear the phase factors $\{e^{i\alpha _{j}}\}_{j=1}^{d}.$ When we
multiply $|\psi \rangle $ by a phase factor $e^{i\alpha _{0}}$ with $\alpha
_{0}\in R$ we get $e^{i\alpha _{0}}|\psi \rangle ,$ and $\alpha _{0}$ is
called a global phase of $e^{i\alpha _{0}}|\psi \rangle .$ Global phase
is nonobservable, then $|\psi \rangle $ and $e^{i\alpha _{0}}|\psi \rangle $
are essentially equivalent. To avoid the freedom of global phase, it is
better to write pure states in the operator form
\begin{eqnarray}
|\psi \rangle \langle \psi |=\sum_{j,k=1}^{d}|c_{j}c_{k}|e^{i(\alpha
_{j}-\alpha _{k})}|j\rangle \langle k|,   \label{Eq1}
\end{eqnarray}%
where, the phase angles $\{\alpha _{j}-\alpha _{k}\}_{j,k=1}^{d}$ are called
relative phases.

A mixed state is a probability combination of some pure states $\rho
=\sum_{\lambda }p_{\lambda }|\psi _{\lambda }\rangle \langle \psi _{\lambda
}|,$ here $\{p_{\lambda }\}_{\lambda}$ is a probability
distribution and $\{|\psi _{\lambda }\rangle \}_{\lambda}$
normalized pure states. Write out $\rho $ in the fixed orthonormal basis $%
\{|j\rangle \}_{j=1}^{d}$ as
\begin{eqnarray}
\rho =\sum_{j,k=1}^{d}|\rho _{jk}|e^{i\alpha _{jk}}|j\rangle \langle k|, \label{Eq2}
\end{eqnarray}%
where $\rho _{jk}=\langle j|\rho |k\rangle =|\rho _{jk}|e^{i\alpha _{jk}}.$
Since each $|\psi _{\lambda }\rangle \langle \psi _{\lambda }|$ in $\rho $
does not contain global phase, then we regard $\rho $ not to contain global
phase, i.e., $\rho $ contains only relative phases. In this letter we only
consider relative phases.

Quantum resource theories provide a unified way to characterize properties
of quantum systems \cite{IJMPB-Horodechi-2013,RMP-Gour-2019}. A well known
quantum resource theory is the entanglement theory
\cite{PRL-Vedral-1997,QIC-Plenio-2007,RMP-Horodecki-2009}. In recent years,
there have been developped other quantum resource theories, such as
coherence \cite{PRL-Plenio-2014,RMP-Plenio-2017}, imaginarity
\cite{JPA-Gour-2018,PRL-Guo-2021,PRA-Guo-2021,Nature-2021-Acin,arXiv-Kondra-2022,arXiv-2023-Guo}, asymmetry \cite{NJP-Gour-2008,Gour-PRA-2009,marvian-Nature-2014}, steering \cite{Gallego-PRX-2015}, contextuality \cite{Amaral-PRL-2018}, and superposition \cite{Plenio-PRL-2017}.

In a quantum resource theory of quantum states, a set of states are
specified as free states which have no certain property, and other states
are called resourceful states. Free operations are defined as operations
which can not create such resource when acting on free states. There are
often different classes of free operations under diverse physical
considerations, such as in coherence theory
\cite{Chitambar-PRL-2016,Chitambar-PRA-2016}. State conversion and resource
measure are important problems in a quantum resource theory.

In coherence theory, the free states (incoherent states) are defined as
diagonal states $\rho =\sum_{j=1}^{d}\rho _{jj}|j\rangle \langle j|$ in
fixed orthonormal basis $\{|j\rangle \}_{j=1}^{d}$. Then coherence theory
quantifies how much the off-diagonal part in states. In imaginarity theory,
the free states (real states) are the states whose entries are all real $%
\rho _{jk}\in R$ in fixed orthonormal basis $\{|j\rangle \}_{j=1}^{d}$. Then
imaginarity theory quantifies how much the imaginary part in states. In this
letter, we ask the question that how much phase contained in quantum states
with respect to the fixed orthonormal basis $\{|j\rangle \}_{j=1}^{d}$. We
will establish a resource theory for the phase of quantum states. To this
aim, we will specify the free states, free operations and propose the
conditions for valid phase measures. In the following we state and discuss our results carefully and put some necessary proofs and details in \cite{SM-Xu-2023}. Note that we always fix the
orthonormal basis $\{|j\rangle \}_{j=1}^{d}$.

\emph{Free states and free operations.}---We first analyze the phase of complex numbers. For a complex number $z=x+iy=|z|e^{i\alpha}$, with $x$ the real part, $y$ the imaginary part, and $\alpha$ the phase angle, $z$ is zero-phase if and only if $z\geq0$. When $\alpha\neq0$, we ask how to quantify the phase contained in $z$? Intuitively, we think that a quantifier of phase should be proportional to $|z|$ and invariant under the transformation $\alpha\rightarrow-\alpha$. These considerations give us some inspirations for the phase of the density operators of quantum states.

We call a state $\rho $ zero-phase if $\langle j|\rho |k\rangle \geq 0$ for
all $j,k.$ In other words, a zero-phase state $\rho $ is a nonnegative state
$\rho \geq 0.$ We denote the set of all zero-phase states by $S_{F}.$
Obviously, $S_{F}$ is a convex set. The convexity of $S_{F}$ leads to significant consequences in both physics and mathematics. In physics, convexity of $S_{F}$ means that a probabilistic mixing of zero-phase states can not create phase, this lays a foundation for the convexity of phase measures (see (P4) in next section). In mathematics, convexity of $S_{F}$ implies that we can employ the techniques of convex optimization.

From Eq. (\ref{Eq1}) we see that a pure zero-phase state can be written in the form $%
|\psi \rangle =\sum_{j=1}^{d}r_{j}|j\rangle $ with $r_{j}\geq 0$, here we
have omitted the global phase. We call such $|\psi \rangle $ a nonnegative
pure state and write it as $|\psi \rangle \geq 0.$

For mixed states, there arises a question that whether a mixed zero-phase
state $\rho \geq 0$ can be decomposed into a probability combination of some
pure zero-phase states. In fact, when $d=2,3,$ any mixed zero-phase mixed
state can be decomposed into a probability combination of pure zero-phase
states \cite{JPA-Xu-2022}.

A quantum operation $\phi $ can be represented by Kraus operators $\phi
=\{K_{\mu }\}_{\mu }$ satisfying $\sum_{\mu }K_{\mu }^{\dagger }K_{\mu
}\preceq I$ with $K_{\mu }^{\dagger }$ the conjugate transpose of $K_{\mu }$ and $I$ the identity operator on $C^{d}.$ Here $\sum_{\mu
}K_{\mu }^{\dagger }K_{\mu }\preceq I$ means that $I-\sum_{\mu
}K_{\mu }^{\dagger }K_{\mu }$ is positive semidefinite. When $\sum_{\mu
}K_{\mu }^{\dagger }K_{\mu }=I,$ the quantum operation $\phi =\{K_{\mu
}\}_{\mu }$ is called a quantum channel \cite{Nielsen-2000-book}. For a quantum
operation $\phi =\{K_{\mu }\}_{\mu },$ if $K_{\mu }\sigma K_{\mu }^{\dagger
}\geq 0$ for all $\mu $ and all zero-phase state $\sigma ,$ we call $\phi
=\{K_{\mu }\}_{\mu }$ a phase nongenerating operation. For the structure of
phase nongenerating operations we have Theorem 1 below.

\emph{Theorem 1.}---A quantum operation $\phi $ is phase nongenerating if and
only if $\phi $ can be written in the Kraus operators $\phi =\{Z_{\lambda
}\}_{\lambda }\cup \{L_{\mu }\}_{\mu },$ where each $Z_{\lambda }$ has the
form $|s\rangle \langle \psi |$ with $|s\rangle $ nonnegative state, and
each $L_{\mu }\geq 0.$

In Theorem 1, we denote $O_{0}=\{\{Z_{\lambda }\}_{\lambda }\},$ and call $%
\{Z_{\lambda }\}_{\lambda }$ a completely phase destroying operation. Such $%
Z_{\lambda }=|s\rangle \langle \psi |$ operates the state $\rho $ into a
pure zero-phase state $|s\rangle \langle \psi |\rho |\psi \rangle \langle s|,
$ completely destroyed the phase of $\rho .$ In Theorem 1, we denote $%
O_{1}=\{\{L_{\mu }\}_{\mu }\},$ and call $\{L_{\mu }\}_{\mu }$ a nonnegative
operation.

In coherence theory, an operation $\phi =\{K_{\mu }\}_{\mu }$ is called
incoherent if each column of $K_{\mu }$ has at most one nonzero entry for
all $\mu $ \cite{PRA-Du-2015}. In contrast, we see that for nonnegative
operation $\{L_{\mu }\}_{\mu }\in O_{1},$ there are too many degrees of
freedom for the entries in each $L_{\mu }$, this fact will make the set $%
O_{1}$ very large and the conditions (P2) and (P3) for phase measures (see next section) very stringent.
With this consideration, we introduce a subset $O_{2}\subset O_{1},$ defined
by $O_{2}=\{\{M_{\mu }\}_{\mu }:$ $\{M_{\mu }\}_{\mu }\in O_{1},\{M_{\mu
}\}_{\mu }$ is incoherent\}. We call $O_{2}$ the set of nonnegative
incoherent operations. $\{K_{\mu }\}_{\mu }\in O_{0}\bigcap O_{1}$ if and only if each $K_{\mu }$ has the form $K_{\mu }=|s\rangle \langle \psi |$ with $|s\rangle \geq 0$ and $|\psi\rangle \geq 0$. $\{K_{\mu }\}_{\mu }\in O_{0}\bigcap O_{2}$ if and only if each $K_{\mu }$ has the form $K_{\mu }=|j\rangle \langle \psi |$ with $|\psi\rangle \geq 0$.
The inclusion relations between $O_{0}$, $O_{1}$, and $O_{2}$ are shown in FIG.1.

\begin{figure}[!ht]
\includegraphics[width=0.4\textwidth,bb=250 110 760 410,clip]{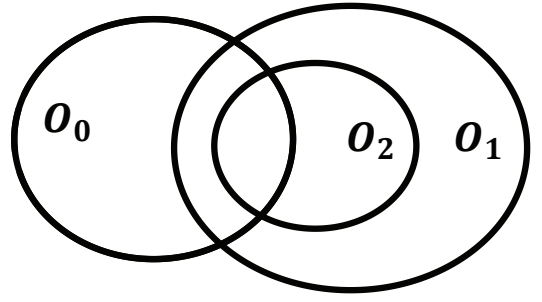}
\caption{Inclusion relations between $O_{0}$, $O_{1}$, and $O_{2}.$}
\end{figure}

For the sake of rigor, we assume the empty set $\emptyset \in O_{0}$ and $\emptyset \in O_{2}\subset O_{1}.$ We use $C_{F}$ to denote the
set of all phase nongenerating channels. Then Theorem 1 says that a channel $\phi
\in C_{F}$ if and only if  $\phi \in O_{0}\cup O_{1}.$ Similarly, we use $%
C_{F}^{\prime }$ to denote the set of all channels such that $\phi \in
O_{0}\cup O_{2}.$ Again $C_{F}^{\prime }\subset C_{F}.$

\emph{Corollary 1.}---A unitary phase nongenerating channel is just a
permutation transformation.

A unitary channel $\phi$ has only one Kraus operator $\phi=\{K\}$ and it satisfies $K^{\dagger}K=I$. By Theorem 1, we see that $\{K\}\in O_{1}$ and $K$ must be a permutation matrix.

Proposition 1 below shows that any operation can be extended to a channel by
adding an appropriate completely phase destroying operation. This fact strengthens the significance of completely phase destroying operations.

\emph{Proposition 1.}---For any operation $\{K_{\mu }\}_{\mu },$ there exists $%
\{Z_{\lambda }\}_{\lambda }\in $ $O_{0}$ such that $\{Z_{\lambda
}\}_{\lambda }\cup \{K_{\mu }\}_{\mu }$ becomes a channel.
As a special case, for any nonnegative operation $\{L_{\mu }\}_{\mu },$
there exists $\{Z_{\lambda }\}_{\lambda }\in $ $O_{0}$ such that $%
\{Z_{\lambda }\}_{\lambda }\cup \{L_{\mu }\}_{\mu }\in C_{F}.$

Since $I-\sum_{\mu }K_{\mu }^{\dagger }K_{\mu }$ is positive
semidefinite, then it has the eigendecomposition $I-\sum_{\mu }K_{\mu
}^{\dagger }K_{\mu }=\sum_{\lambda }s_{\lambda }^{2}|\psi _{\lambda }\rangle
\langle \psi _{\lambda }|$ with $s_{\lambda }>0.$ There obviously exist $|s_{\lambda
}\rangle \geq 0$ satisfying $\langle s_{\lambda }|s_{\lambda }\rangle
=s_{\lambda }^{2}.$ Let $Z_{\lambda }=|s_{\lambda }\rangle \langle \psi _{\lambda
}|$, then $\{Z_{\lambda }\}_{\lambda }\in $ $O_{0}$ and $\{Z_{\lambda }\}_{\lambda }\cup \{K_{\mu
}\}_{\mu }$ becomes a channel.

\emph{A framework for quantifying the phase of quantum states under $C_{F}$}---A measure of the phase for quantum states is a real-valued functional
$\mathcal{P}(\rho )$ satisfying some conditions. Inspired by the coherence theory \cite{PRL-Plenio-2014}, we
propose the following conditions (P1) to (P4) which any phase measure should satisfy
under $C_{F}$.

(P1). Faithfulness. $\mathcal{P}(\rho )\geq 0$ for any state $\rho $ and $\mathcal{P}(\rho )=0$
if and only if $\rho $ is zero-phase.

(P2). Monotonicity. $\mathcal{P}(\sum_{\mu }K_{\mu }\rho K_{\mu }^{\dagger })\leq
\mathcal{P}(\rho )$ for any state $\rho $ and any $\{K_{\mu }\}_{\mu }\in C_{F}.$

(P3). Probabilistic monotonicity. $\sum_{\mu }[$tr$(K_{\mu }\rho K_{\mu
}^{\dagger })]\mathcal{P}(\frac{K_{\mu }\rho K_{\mu }^{\dagger }}{\text{tr}(K_{\mu
}\rho K_{\mu }^{\dagger })})\leq \mathcal{P}(\rho )$ for any state $\rho $ and any $%
\{K_{\mu }\}_{\mu }\in C_{F}.$

(P4). Convexity. $\mathcal{P}(\sum_{\mu }p_{\mu }\rho _{\mu })\leq \sum_{\mu }p_{\mu
}\mathcal{P}(\rho _{\mu })$ for any states $\{\rho _{\mu }\}_{\mu }$ and any
probability distribution $\{p_{\mu }\}_{\mu }.$

Similar to the coherence theory, one can prove that (P3) and (P4) imply
(P2). Further, (P3) and (P4) imply (P5) below. (P5) also holds for coherence measures in coherence theory \cite{PRA-Tong-2016}.

(P5). Additivity for direct sum states.
\begin{eqnarray}
\mathcal{P}(p\rho _{1}\oplus (1-p)\rho _{2})= p\mathcal{P}(\rho _{1})+(1-p)\mathcal{P}(\rho _{2}) \label{Eq3}
\end{eqnarray}%
for states $\rho _{1},\rho _{2}$ and $p\in (0,1).$

From Corollary 1, (P2) and (P3), we see that for any phase measure $\mathcal{P}(\rho
),$ a permutation on any state preserves the phase of this state. Hence, in Eq. (\ref{Eq3}), without loss of generality, we assume that there exists positive integer $d_{1}$ such that $\sum_{j=1}^{d_{1}}|j\rangle\langle j|\rho\sum_{k=1}^{d_{1}}|k\rangle\langle k|=p\rho _{1}$ and $\sum_{j=d_{1}+1}^{d}|j\rangle\langle j|\rho\sum_{k=d_{1}+1}^{d}|k\rangle\langle k|=(1-p)\rho _{2}$. Thus (P3) yields $\mathcal{P}(p\rho _{1}\oplus (1-p)\rho _{2})\geq p\mathcal{P}(\rho _{1})+(1-p)\mathcal{P}(\rho _{2})$. Further, since $p\rho _{1}\oplus (1-p)\rho _{2}=p\rho _{1}+(1-p)\rho _{2}$, thus (P4) yields $\mathcal{P}(p\rho _{1}\oplus (1-p)\rho _{2})\leq p\mathcal{P}(\rho _{1})+(1-p)\mathcal{P}(\rho _{2}).$
Combine these two inequalities, then (P5) follows.

We introduce the property of conjugation invariance for phase measures. A phase measure $\mathcal{P}(\rho)$ is called conjugation invariant or $\mathcal{P}(\rho)$ has the property of conjugation invariance if
\begin{eqnarray}
\mathcal{P}(\rho )=\mathcal{P}(\rho^{*})    \label{Eq4}
\end{eqnarray}
for any state $\rho$. Where $\rho^{*}$ is the conjugate state of $\rho$, that is, $\langle j|\rho^{*}|k\rangle=(\langle j|\rho|k\rangle)^{*}$ with $^{*}$ denoting the complex conjugation.

The definition of robustness is used in many quantum resource theories. It gives rise to an
entanglement measure in entanglement theory \cite{QIC-Plenio-2007,PRA-Vidal-1999,PRA-Steiner-2003}, a coherence measure in coherence theory \cite{PRL-Adesso-2016,PRA-Adesso-2016} and an imaginarity measure in imaginarity theory \cite{JPA-Gour-2018,PRL-Guo-2021,PRA-Guo-2021}. For the phase of quantum states, we prove that robustness also yields a valid phase measure.

\emph{Theorem 2.}---Robustness of phase defined by
\begin{eqnarray}
\mathcal{P}_{\text{rob}}(\rho )=\min_{\tau }\{s:\rho +s\tau \geq 0\} \label{Eq5}
\end{eqnarray}
is a phase measure under $C_{F},$ where min runs over all quantum states.

$\mathcal{P}_{\text{rob}}(\rho )$ is conjugation invariant, this is the consequence of that if $\rho +s\tau \geq 0$ then $\rho^{*} +s\tau^{*} \geq 0$. Also, $\mathcal{P}_{\text{rob}}(\rho )$ has closed expressions for any qubit state, this is Theorem 3 below.

\emph{Theorem 3.}--- For the qubit state $\rho $, we express $\rho $ in Bloch representation as
\begin{eqnarray}
\rho =\frac{1}{2}\left(
\begin{array}{cc}
1+z & x-iy \\
x+iy & 1-z%
\end{array}%
\right) ,  \label{III.1}
\end{eqnarray}
where $(x,y,z)$ is a real vector called Bloch vector satisfying $%
x^{2}+y^{2}+z^{2}\leq 1.$
Then the robustness of phase for $\rho $ is
\begin{eqnarray}
\mathcal{P}_{\text{rob}}(\rho )=
   \begin{cases}
\ \ \ \ \ \ |y|, & x\geq 0, \\
\sqrt{x^{2}+y^{2}}, & x< 0.
   \end{cases}   \label{III.12}
\end{eqnarray}

\emph{Phase measures under $C_{F}^{\prime }$}---In conditions (P2) and (P3), if we replace $C_{F}$ by $C_{F}^{\prime }$ we
will obtain conditions of phase measures under $C_{F}^{\prime }.$ Since $C_{F}^{\prime }\subset C_{F},$ then any phase measure under $C_{F}$ is
automatically a phase measure under $C_{F}^{\prime },$ such as $\mathcal{P}_{\text{rob}%
}(\rho )$. Similarly to the case of $C_{F}$, we can prove that under $C_{F}^{\prime }$, (P3) and (P4) imply (P5).
Now we provide a phase measure $%
\mathcal{P}_{1}(\rho )$ under $C_{F}^{\prime }.$

\emph{Theorem 4.}---For quantum states $\rho$,
\begin{eqnarray}
\mathcal{P}_{1}(\rho )=\sum_{j\neq k}(|\rho _{jk}|-\text{Re}\rho_{jk})   \label{Eq7}
\end{eqnarray}
is a valid phase measure under $C_{F}^{\prime },$ where Re$\rho_{jk}$ is the real part of $\rho_{jk}$.

Different from $\mathcal{P}_{\text{rob}}(\rho )$, $\mathcal{P}_{1}(\rho )$ is the sum $\mathcal{P}_{1}(\rho )=\sum_{j,k}\mathcal{P}_{1}(\rho_{jk} )$ for all entries $\rho_{jk}$ with $\mathcal{P}_{1}(\rho_{jk} )=|\rho _{jk}|-\text{Re}\rho_{jk}$. Notice that $\mathcal{P}_{1}(\rho )=\sum_{j\neq k}\mathcal{P}_{1}(\rho_{jk} )=\sum_{j,k}\mathcal{P}_{1}(\rho_{jk} )$ since $\mathcal{P}_{1}(\rho_{jj})=0$. $\mathcal{P}_{1}(\rho )$ obviously satisfies the additivity for direct sum states in Eq. (\ref{Eq3}) and the conjugation invariance in Eq. (\ref{Eq4}).

\emph{Example 1.}---Phase of the coherent phase states. The coherent phase
states are defined as \cite{PPF-Susskind-1964,PRA-Shapiro-1991}
\begin{eqnarray}
|\varepsilon \rangle =\sqrt{1-|\varepsilon |^{2}}\sum_{j=0}^{\infty
}\varepsilon ^{j}|j\rangle ,\varepsilon \in C,|\varepsilon |<1.   \label{Eq8}
\end{eqnarray}
We can directly derive $\mathcal{P}_{1}(|\varepsilon \rangle )$ as
\begin{eqnarray}
\mathcal{P}_{1}(|\varepsilon \rangle )=(1-|\varepsilon |^{2})[\frac{1}{%
(1-|\varepsilon |)^{2}}-\frac{1}{|1-\varepsilon |^{2}}].    \label{Eq8}
\end{eqnarray}

\emph{Theorem 5.}---For $\mathcal{P}_{1}(\rho ),$ it holds that
\begin{eqnarray}
\mathcal{P}_{1}(\rho )\leq d.   \label{Eq10}
\end{eqnarray}
Also, a pure state reaching the maximum of $\mathcal{P}_{1}(\rho )$, $\mathcal{P}_{1}(\rho )=d$, must be of the form
\begin{eqnarray}
|\psi _{\max }(\theta )\rangle =\frac{1}{\sqrt{d}}\sum_{j=1}^{d}e^{i\theta
_{j}}|j\rangle ,\sum_{j=1}^{d}e^{i\theta _{j}}=0.   \label{Eq11}
\end{eqnarray}

Because of (P4), there must exist
pure state reaching the maximum of $\mathcal{P}_{1}(\rho ).$ For a normalized pure
state $|\psi \rangle =\sum_{j=1}^{d}c_{j}|j\rangle $ with $%
\sum_{j=1}^{d}|c_{j}|^{2}=1,$ we have $\mathcal{P}_{1}(|\psi \rangle
)=(\sum_{j}|c_{j}|)^{2}-|\sum_{j}c_{j}|^{2}.$ To maximize $\mathcal{P}_{1}(|\psi
\rangle ),$ we let $\sum_{j}c_{j}=0,$ and $\sum_{j}|c_{j}|^{2}$ be maximal.
Applying the Lagrange multiplier method, one can prove that under the
condition $\sum_{j=1}^{d}|c_{j}|^{2}=1,$ $\sum_{j}|c_{j}|^{2}$ is maximal if
and only if $|c_{j}|=\frac{1}{\sqrt{d}}$ for any $j.$ Then Theorem 5
follows.

In Eq. (\ref{Eq11}), there is a special state that $\theta _{j}=(j-1)\frac{2\pi }{d}$
and
\begin{eqnarray}
|\psi _{\max }\rangle =\frac{1}{\sqrt{d}}\sum_{j=1}^{d}e^{i(j-1)\frac{2\pi }{%
d}}|j\rangle .   \label{Eq12}
\end{eqnarray}

Since $\mathcal{P}_{1}(|\psi \rangle )$ is preserved under any permutation on $|\psi
\rangle $, and $\mathcal{P}_{1}(|\psi \rangle )$ is preserved by adding any global
phase factor, then with some permutations and an appropriate global phase factor, we can get $\theta =\{\theta
_{j}\}_{j=1}^{d}$ in standard form as $\theta ^{\uparrow }=\{\theta
_{j}^{\uparrow }\}_{j=1}^{d}$: $0=\theta _{1}^{\uparrow }\leq \theta
_{2}^{\uparrow }\leq ...\leq \theta _{d}^{\uparrow }<2\pi .$ We represent $%
|\psi _{\max }(\theta ^{\uparrow })\rangle $ in complex plane in FIG.2 as follows: $%
|\psi _{\max }(\theta ^{\uparrow })\rangle $ is represented by a convex
equilateral polygon with vertices $\{M_{j}\}_{j=0}^{d},$ where $M_{d}=M_{0}$
is the origin of the plane, $\overrightarrow{M_{j-1}M_{j}}=\frac{1}{\sqrt{d}}%
e^{i\theta _{j}^{\uparrow }}.$ In particular, $|\psi _{\max }\rangle $ is
represented by a regular (equilateral and equiangular) polygon with $%
\overrightarrow{M_{j-1}M_{j}}=\frac{1}{\sqrt{d}}e^{i(j-1)\frac{2\pi }{d}}.$
From this representation, we see that for $d=2,$ there is only one pure
state with standard form $\{\theta _{1}^{\uparrow }=0,\theta _{2}^{\uparrow
}=\pi \}$ which reaches the maximum of $\mathcal{P}_{1}(\rho ).$ For $d=3,$ there is
only one pure state with standard form $\{\theta _{1}^{\uparrow }=0,\theta
_{2}^{\uparrow }=\frac{2\pi }{3},\theta _{2}^{\uparrow }=\frac{4\pi }{3}\}$
which reaches the maximum of $\mathcal{P}_{1}(\rho ),$ see FIG.3. For $d\geq 4,$ there is an
infinite number of pure states with standard form $\theta ^{\uparrow
}=\{\theta _{j}^{\uparrow }\}_{j=1}^{d}$ which reaches the maximum of $%
\mathcal{P}_{1}(\rho ).$

\begin{figure}[!ht]
\includegraphics[width=0.45\textwidth,bb=230 80 730 520,clip]{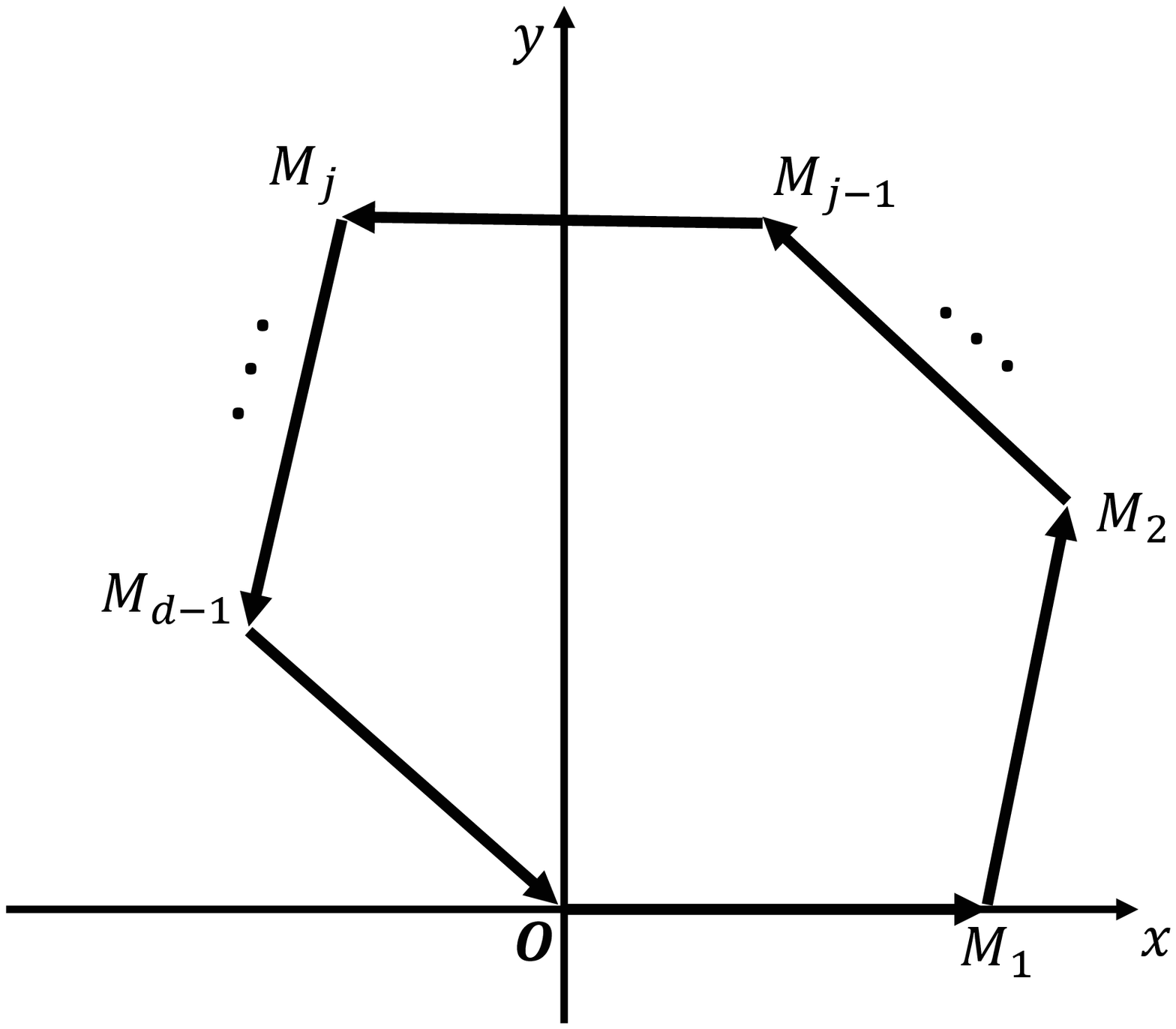}
\caption{$|\psi _{\max }(\theta)\rangle $ with standard form in Eq. (\ref{Eq12}) is represented by a convex equilateral polygon in complex plane.}
\includegraphics[width=0.35\textwidth,bb=320 80 740 490,clip]{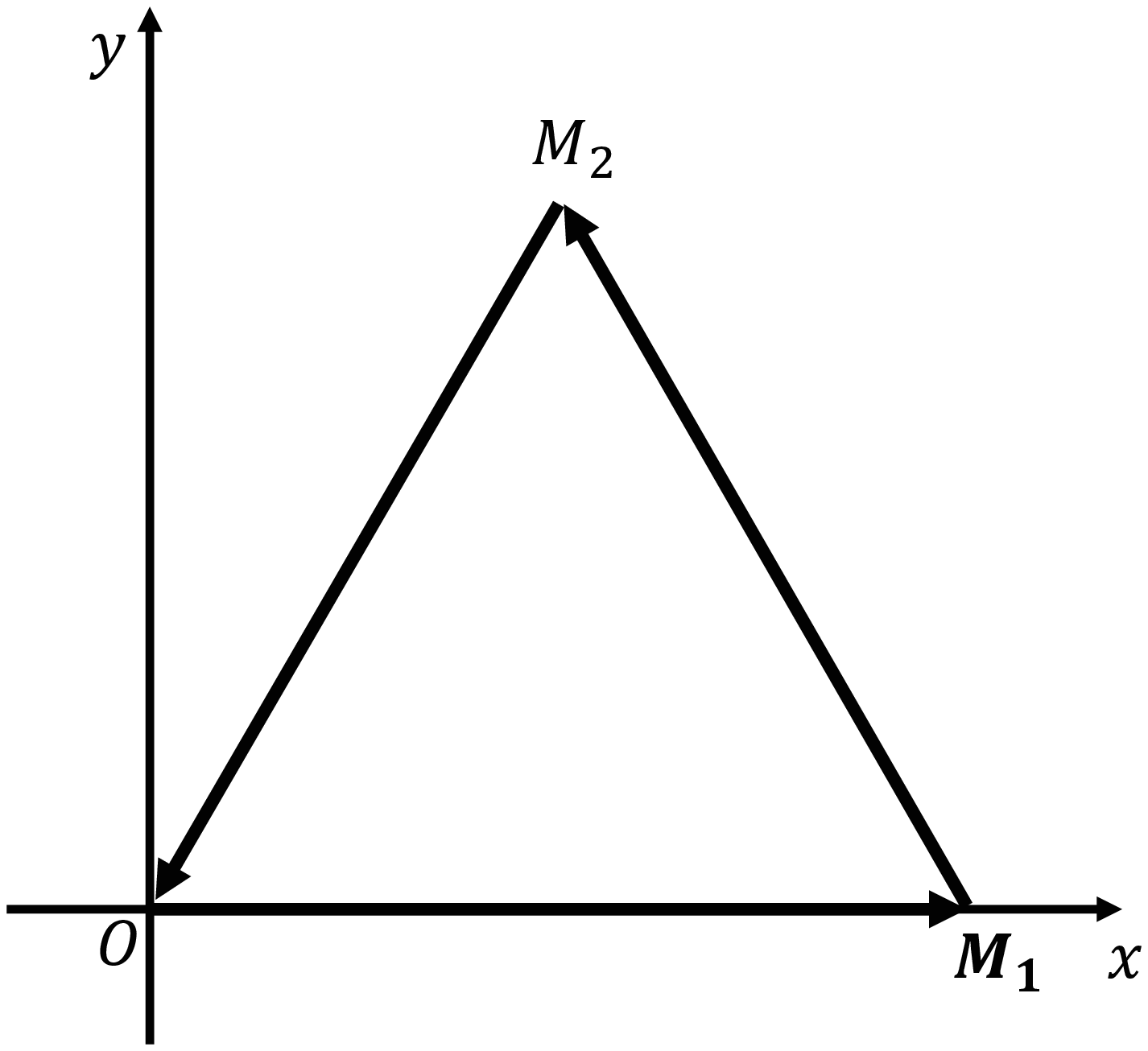}
\caption{For $d=3,$ there is only one $|\psi _{\max }(\theta)\rangle $ with standard form in Eq. (\ref{Eq12}) which is represented by an equilateral triangle in complex plane.}
\end{figure}

\emph{State conversion}---State conversion is one of the main problems in a resource theory. Here we
provide two results for state conversion in resource theory of phase.

\emph{Proposition 2.}---Any state can be converted into a fixed pure
zero-phase state via a channel $\{Z_{\lambda }\}_{\lambda }\in O_{0}.$

For fixed pure zero-phase state $|\varphi \rangle
=\sum_{j=1}^{d}r_{j}|j\rangle $ with $r_{j}\geq 0$ and $%
\sum_{j=1}^{d}r_{j}^{2}=1,$
let $Z_{j}=|\varphi \rangle \langle \psi _{j}|$ with $\{|\psi _{j}\rangle
\}_{j=1}^{d}$ orthonormal vectors. Then $\{Z_{j}\}_{j=1}^{d}$ is a channel
and $\{Z_{j}\}_{j=1}^{d}\in O_{0}.$
Further, for any state $\rho $, we have $Z_{j}\rho Z_{j}^{\dagger }=|\varphi
\rangle \langle \psi _{j}|\rho |\psi _{j}\rangle \langle \varphi |$ and $%
\sum_{j=1}^{d}Z_{j}\rho Z_{j}^{\dagger }=|\varphi \rangle \langle \varphi |.$

\emph{Proposition 3.}---For pure state $|\varphi \rangle
=\sum_{j=1}^{d}|c_{j}|e^{i(j-1)\frac{2\pi }{d}}|j\rangle $ with $%
\sum_{j=1}^{d}|c_{j}|^{2}=1$, and $|\psi _{\max }\rangle $ in Eq. (\ref{Eq12}), there
exists a channel $\{M_{\mu}\}_{\mu}\in O_{2}$ which converts $|\psi _{\max
}\rangle $ to $|\varphi \rangle .$

Let $M_{j}=\sum_{k=1}^{d}|c_{k}||k\rangle \langle k+j|$ with $|k+j\rangle=|(k+j) \ \text{mod} \ d\rangle,$ then $%
\{M_{j}\}_{j=1}^{d}$ is a channel and $\{M_{j}\}_{j=1}^{d}\in O_{2}.$
In addition, one gets that $M_{j}|\psi _{\max }\rangle =\frac{1}{\sqrt{d}}%
e^{ij\frac{2\pi }{d}}|\varphi \rangle $ and $\sum_{j=1}^{d}M_{j}|\psi _{\max
}\rangle \langle \psi _{\max }|M_{j}^{\dagger }=|\varphi \rangle \langle
\varphi |.$

\emph{Intrinsic phase of quantum states.}---For a phase measure $\mathcal{P}(\rho )$ under $C_{F}$ or $C_{F}^{\prime }$, we define the intrinsic phase as
\begin{eqnarray}
\overline{\mathcal{P}}(\rho )=\min_{\{\theta _{j}\}_{j=1}^{d}\subset R}\mathcal{P}(U_{\theta
}\rho U_{\theta }^{\dagger }),    \label{Eq13}
\end{eqnarray}
where $U_{\theta }=$diag$(e^{i\theta _{1}},e^{i\theta _{2}},...,e^{i\theta
_{d}}).$

If $\mathcal{P}$ is conjugation invariant, then so is $\overline{\mathcal{P}}(\rho )$.  $\overline{\mathcal{P}}(\rho )=0$ if and only if there exist $\{\theta
_{j}\}_{j=1}^{d}\subset R$ such that
\begin{eqnarray}
\rho =\sum_{j,k=1}^{d}|\rho _{jk}|e^{i(\theta _{j}-\theta _{k})}.   \label{Eq14}
\end{eqnarray}%
If $\overline{\mathcal{P}}(\rho )=0,$ we call $\rho $ an intrinsically zero-phase
state. From Eq. (\ref{Eq14}) we see that $\overline{\mathcal{P}}(\rho )=0$ for pure states,
zero-phase states and qubit states. For $d=3$, there exist some states that $\overline{\mathcal{P}}(\rho )>0$ \cite{JPA-Xu-2022}.

For the fixed orthonormal basis $\{|j\rangle \}_{j=1}^{d}$ of $C^{d},$ $%
\{e^{i\theta _{j}}|j\rangle \}_{j=1}^{d}$ is also an orthonormal basis of $%
C^{d}.$ Then $\overline{\mathcal{P}}(\rho )$ can be viewed as the minimum phase $\mathcal{P}(\rho )$ with respect to all $\{e^{i\theta _{j}}|j\rangle \}_{j=1}^{d}$ for $\{\theta _{j}\}_{j=1}^{d}\subset R$.

\emph{Summary and outlook.}---We introduced a resource theory for the phase of quantum states. To do this,
we specified the free states and two kinds of free channels. We proposed
some conditions for phase measures, and provided two such phase
measures. We also introduced the concept of intrinsic phase.

There remained many open questions for future research. First, the
conditions for state conversions need further in-depth investigations,
including deterministic conversion, probabilistic conversion and catalytic
conversion, as detailed discussions in entanglement theory and in coherence theory. Second, more phase measures and their possible operational
interpretations in relevant experimental scenarios are very desirable, these will deepen the understanding and facilitate the applications of phase theory introduced in this letter.
Third, the relationships between conjugation invariance in Eq. (\ref{Eq4}) and (P1) to (P5) are worthy of attention.
\\
\\

\section{}
\textbf{Supplemental Material: Quantifying the phase of quantum states} \\
\text{\ \ \ \ \ \ \ \ \ \ \ \ \ \ \ \ \ \ \ \ \ \ \ \  Jianwei Xu}

\emph{College of Science, Northwest A$\&$F University, Yangling, Shaanxi 712100, China}

\section{Proof of Theorem 1: Structure of phase nongenerating operations}
\setcounter{equation}{0} \renewcommand\theequation{I.\arabic{equation}}

For clarity, we set three steps for this proof.

(1). Suppose $\phi =\{K_{\mu }\}_{\mu }$ is a phase nongenerating operation,
then $K_{\mu }\sigma K_{\mu }^{\dagger }\geq 0$ for any state $\sigma \geq 0$
and any $\mu .$ Expand $K_{\mu }$ as $K_{\mu }=\sum_{j,k=1}^{d}K_{\mu
,jk}|j\rangle \langle k|$ with $K_{\mu ,jk}=\langle j|K_{\mu }|k\rangle .$
For the zero-phase state
\begin{eqnarray}
\sigma =p|l\rangle \langle l|+(1-p)|m\rangle \langle m|+x|l\rangle \langle
m|+x|m\rangle \langle l|  \label{I.1}
\end{eqnarray}%
with $1\leq l\neq m\leq d$, $p\in \lbrack 0,1]$, $x\in \lbrack 0,\sqrt{p(1-p)%
}]$, then
\begin{eqnarray}
&&K_{\mu }\sigma K_{\mu }^{\dagger }  \nonumber \\
&=&\sum_{j,j^{\prime }=1}^{d}[pK_{\mu
,jl}K_{\mu ,j^{\prime }l}^{\ast }+(1-p)K_{\mu ,jm}K_{\mu ,j^{\prime
}m}^{\ast }  \nonumber \\
&&\ \ \ \ \ \ +x(K_{\mu ,jl}K_{\mu ,j^{\prime }m}^{\ast }+K_{\mu ,jm}K_{\mu ,j^{\prime
}l}^{\ast })]|j\rangle \langle j^{\prime }|.   \label{I.2}
\end{eqnarray}

Let $p=x=0,$ then $K_{\mu }\sigma K_{\mu }^{\dagger }\geq 0$ yields
\begin{eqnarray}
&&K_{\mu ,jm}K_{\mu ,j^{\prime }m}^{\ast } \geq 0\text{ }\forall j,j^{\prime
},m,  \label{I.3} \\
&&K_{\mu ,jm} =|K_{\mu ,jm}|e^{i\theta _{m}},\theta _{m}\in R,\forall
j,j^{\prime },m. \label{I.4}
\end{eqnarray}

We see that for fixed $l$ if $K_{\mu ,jl}=0$ for all $j,$  i.e., the $l$-th
column is a zero column, then Eq. (\ref{I.3}) always holds. Hence below we
exclude such zero columns.

Let $p\in (0,1)$ and $x=\sqrt{p(1-p)},$ then $K_{\mu }\sigma K_{\mu
}^{\dagger }\geq 0$ yields
\begin{eqnarray}
&&\ \ p|K_{\mu ,jl}K_{\mu ,j^{\prime }l}|+(1-p)|K_{\mu ,jm}K_{\mu ,j^{\prime }m}|
\nonumber \\
&&+\sqrt{p(1-p)}[|K_{\mu ,jl}K_{\mu ,j^{\prime }m}|e^{i(\theta _{l}-\theta
_{m})} \nonumber \\
&& +|K_{\mu ,jm}K_{\mu ,j^{\prime }l}|e^{-i(\theta _{l}-\theta _{m})}]
\geq 0,  \label{I.5} \\
&&\ \ (|K_{\mu ,jl}K_{\mu ,j^{\prime }m}|-|K_{\mu ,jm}K_{\mu ,j^{\prime }l}|)\sin
(\theta _{l}-\theta _{m}) =0. \ \ \ \ \ \  \label{I.6}
\end{eqnarray}

Eq. (\ref{I.6}) implies that
\begin{equation}
\theta _{l}-\theta _{m}=n\pi \ (\text{mod} \ 2\pi), n\in \{0,1\}, \label{I.7}
\end{equation}
or
\begin{equation}
|K_{\mu ,jl}K_{\mu ,j^{\prime }m}|=|K_{\mu ,jm}K_{\mu ,j^{\prime
}l}|,\forall j,j^{\prime }. \label{I.8}
\end{equation}

(2). Suppose the $l$-th and $m$-th columns satisfy $\theta _{l}-\theta
_{m}=\pi ,$ we prove that Eq. (\ref{I.8}) holds. For simplicity of symbols, we let $|K_{\mu ,jl}|=a_{1},$ $|K_{\mu ,j^{\prime
}l}|=a_{2},$ $|K_{\mu ,jm}|=a_{3},$ $|K_{\mu ,j^{\prime }m}|=a_{4},$ thus Eq. (\ref{I.5}) yields
\begin{eqnarray}
pa_{1}a_{2}+(1-p)a_{3}a_{4}-\sqrt{p(1-p)}(a_{1}a_{4}+a_{2}a_{3}) \geq 0.  \ \label{I.9} \\
(\sqrt{p}a_{1}-\sqrt{1-p}a_{3})(\sqrt{p}a_{2}-\sqrt{1-p}a_{4}) \geq 0.  \ \ \ \ \ \  \label{I.10}
\end{eqnarray}

Consider the case that more than one of $\{a_{1},a_{2},a_{3},a_{4}\}$ are zero. If $a_{1}=a_{2}=0$ or $a_{3}=a_{4}=0$ or $a_{1}=a_{3}=0$ or $a_{2}=a_{4}=0$, we see that Eq. (\ref{I.8}) holds. If $a_{1}=a_{4}=0$, Eq. (\ref{I.10}) yields $a_{2}a_{3}=0$, then $a_{2}=0$ or $a_{3}=0$, thus Eq. (\ref{I.8}) holds. Similarly, if $a_{2}=a_{3}=0$, then Eq. (\ref{I.8}) holds.

When there is at most one of $\{a_{1},a_{2},a_{3},a_{4}\}$ being zero, without loss of generality, suppose $a_{3}a_{4}\neq0$, then Eq. (\ref{I.10}) leads to
\begin{eqnarray}
(\frac{a_{1}}{a_{3}}-\sqrt{\frac{1-p}{p}})(\frac{a_{2}}{a_{4}}-\sqrt{\frac{%
1-p}{p}}) \geq 0.  \label{I.11}
\end{eqnarray}
The solutions of Eq. (\ref{I.11}) are
\begin{equation}
\frac{a_{1}}{a_{3}}\geq \sqrt{\frac{1-p}{p}},\frac{a_{2}}{a_{4}}\geq \sqrt{%
\frac{1-p}{p}}, \label{I.12}
\end{equation}
and
\begin{equation}
\frac{a_{1}}{a_{3}}\leq \sqrt{\frac{1-p}{p}},\frac{a_{2}}{a_{4}}\leq \sqrt{%
\frac{1-p}{p}}. \label{I.13}
\end{equation}
$p\in (0,1),$ then $\sqrt{\frac{1-p}{p}}\in (0,+\infty ).$ For given
nonnegative numbers $\{a_{1},a_{2},a_{3},a_{4}\},$ Eq. (\ref{I.12}) or Eq. (\ref{I.13}) holds
for all $p\in (0,1),$ then there must hold that $\frac{a_{1}}{a_{3}}=\frac{%
a_{2}}{a_{4}},$ this is just Eq. (\ref{I.8}). It follows that, for any two columns,
the $l$-th and $m$-th columns, either $\theta _{l}-\theta _{m}=0$ or Eq. (\ref{I.8})
holds.

(3). In set theory, an equivalence relation in a set is a binary relation
that is reflexive, symmetric and transitive. An equivalence relation
provides a partition of the set into disjoint equivalence classes. We can
check that $\theta _{l}-\theta _{m}=0$ and Eq. (\ref{I.8}) are all equivalence
relations of the set of nonzero columns.

If $\theta _{l}-\theta _{m}=0$ for all $l$-th and $m$-th columns, we can
write $K_{\mu ,jl}=|K_{\mu ,jl}|e^{i\alpha }$ for all $j,l,$ and $\alpha \in
R$ is independent of $j,k.$ We can choose $\alpha =0$ since $K_{\mu }\sigma
K_{\mu }^{\dagger }=(e^{i\alpha }K_{\mu })\sigma (e^{i\alpha }K_{\mu
})^{\dag }.$ That is, $K_{\mu }\geq 0.$

Now suppose there are at least two disjoint equivalence classes $S_{1}\cup
S_{2}...$ under the equivalence relation $\theta _{l}-\theta _{m}=0$ for nonzero columns. Suppose the
$l$-th column is contained in $S_{1},$ the $m$-th column is contained in $%
S_{2},$ then the $l$-th and $m$-th columns must satisfy Eq. (\ref{I.8}). Since Eq. (\ref{I.8}) is
also an equivalence relation, then any two nonzero columns of $K_{\mu }$ satisfy Eq. (\ref{I.8}). As a
result, $K_{\mu }$ has the form
\begin{eqnarray}
K_{\mu }=\left(
\begin{array}{cccc}
s_{1}t_{1}e^{i\theta _{1}} & s_{1}t_{2}e^{i\theta _{2}} & ... &
s_{1}t_{d}e^{i\theta _{d}} \\
s_{2}t_{1}e^{i\theta _{1}} & s_{2}t_{2}e^{i\theta _{2}} & ... &
s_{2}t_{d}e^{i\theta _{d}} \\
... & ... & ... & ... \\
s_{d}t_{1}e^{i\theta _{1}} & s_{d}t_{2}e^{i\theta _{2}} & ... &
s_{d}t_{d}e^{i\theta _{d}}%
\end{array}%
\right) ,   \label{I.14}
\end{eqnarray}
with $\{s_{j}\geq 0\}_{j=1}^{d},$ $\{t_{j}\geq 0\}_{j=1}^{d},$ $\{\theta
_{j}\}_{j=1}^{d}\subset R.$ $K_{\mu }$ can be written in a
succinct form as
\begin{equation}
K_{\mu }=|s\rangle \langle \psi |,  \label{I.15}
\end{equation}
where $|s\rangle =\sum_{j=1}^{d}s_{j}|j\rangle \geq 0,$ $|\psi \rangle
=\sum_{j=1}^{d}t_{j}e^{i\theta _{j}}|j\rangle,$ $\langle \psi |\psi
\rangle =1.$

\section{Proof of Theorem 2: Robustness of phase is a phase measure under $%
C_{F}$}
\setcounter{equation}{0} \renewcommand\theequation{II.\arabic{equation}}

(P1) evidently holds.

To prove (P4), let $p\in (0,1)$ and $\rho _{1},$ $%
\rho _{2}$ be any two states. Suppose $\mathcal{P}_{\text{rob}}(\rho _{1})=s_{1}$ and
state $\tau _{1}$ reaches $\mathcal{P}_{\text{rob}}(\rho _{1})=s_{1},$ that is
\begin{eqnarray}
\rho _{1}+s_{1}\tau _{1}\geq 0. \label{II.1}
\end{eqnarray}
Similarly, suppose $\mathcal{P}_{\text{rob}}(\rho _{2})=s_{2}$ and state $\tau _{2}$
reaches $\mathcal{P}_{\text{rob}}(\rho _{2})=s_{2},$ that is
\begin{eqnarray}
\rho _{2}+s_{2}\tau _{2}\geq 0. \label{II.2}
\end{eqnarray}

Eqs. (\ref{II.1},\ref{II.2}) imply
\begin{eqnarray}
p(\rho _{1}+s_{1}\tau _{1})+(1-p)(\rho _{2}+s_{2}\tau _{2}) \geq 0,  \label{II.3} \\
\lbrack p\rho _{1}+(1-p)\rho _{2}]+ps_{1}\tau _{1}+(1-p)s_{2}\tau _{2} \geq 0.  \label{II.4}
\end{eqnarray}
By the definition of $\mathcal{P}_{\text{rob}}$ and Eq. (\ref{II.4}), we get that
\begin{eqnarray}
&&\mathcal{P}_{\text{rob}}(p\rho _{1}+(1-p)\rho _{2}) \nonumber \\
&\leq &\text{tr}[ps_{1}\tau _{1}+(1-p)s_{2}\tau _{2}]=ps_{1}+(1-p)s_{2},  \label{II.5}
\end{eqnarray}
this proves (P4).

Next we prove (P3). Suppose the channel $%
\phi \in C_{F}.$ Employing Proposition 1, $\phi $ can be
expressed as $\phi =\{Z_{\lambda }\}_{\lambda }\cup \{L_{\mu }\}_{\mu }$
with $\{Z_{\lambda }\}_{\lambda }\in O_{0}$ and $\{L_{\mu }\}_{\mu }\in
O_{1}.$ Since $Z_{\lambda }\rho Z_{\lambda }^{\dagger }\geq 0$ and $\mathcal{P}_{\text{%
rob}}(\frac{Z_{\lambda }\rho Z_{\lambda }^{\dagger }}{\text{tr}(Z_{\lambda
}\rho Z_{\lambda }^{\dagger })})=0$ for any $\lambda ,$ then we only need to
prove that
\begin{eqnarray}
\sum_{\mu }[\text{tr}(L_{\mu }\rho L_{\mu }^{t})]\mathcal{P}_{\text{rob}}(\frac{L_{\mu
}\rho L_{\mu }^{t}}{\text{tr}(L_{\mu }\rho L_{\mu }^{t})})\leq \mathcal{P}_{\text{rob}%
}(\rho )
\end{eqnarray}
where $L_{\mu }^{t}$ denotes the transpose of $L_{\mu }.$

Suppose $\mathcal{P}_{\text{rob}}(\rho )=s$ and state $\tau $ reaches $\mathcal{P}_{\text{rob}%
}(\rho )=s,$ that is
\begin{eqnarray}
\rho +s\tau \geq 0.
\end{eqnarray}
It follows that
\begin{eqnarray}
&&L_{\mu }\rho L_{\mu }^{t}+sL_{\mu }\tau L_{\mu }^{t} \geq 0, \nonumber \\
&&\frac{L_{\mu }\rho L_{\mu }^{t}}{\text{tr}(L_{\mu }\rho L_{\mu }^{t})}+\frac{%
s}{\text{tr}(L_{\mu }\rho L_{\mu }^{t})}L_{\mu }\tau L_{\mu }^{t} \geq 0,
 \nonumber \\
&&\mathcal{P}_{\text{rob}}(\frac{L_{\mu }\rho L_{\mu }^{t}}{\text{tr}(L_{\mu }\rho
L_{\mu }^{t})}) \leq \frac{s}{\text{tr}(L_{\mu }\rho L_{\mu }^{t})}\text{tr%
}(L_{\mu }\tau L_{\mu }^{t}),  \nonumber \\
&&\sum_{\mu }[\text{tr}(L_{\mu }\rho L_{\mu }^{t})]\mathcal{P}_{\text{rob}}(\frac{L_{\mu
}\rho L_{\mu }^{t}}{\text{tr}(L_{\mu }\rho L_{\mu }^{t})})  \nonumber  \\
&\leq&s\sum_{\mu
}\text{tr}(L_{\mu }\tau L_{\mu }^{t})\leq s,   \nonumber
\end{eqnarray}
then (P3) holds.

\section{Proof of Theorem 3: Robustness of phase for qubit states}
\setcounter{equation}{0} \renewcommand\theequation{III.\arabic{equation}}
For any qubit state $\rho ,$ we write $\rho $ in the Bloch representation as
\begin{eqnarray}
\rho =\frac{1}{2}\left(
\begin{array}{cc}
1+z & x-iy \\
x+iy & 1-z%
\end{array}%
\right) ,  \label{III.1}
\end{eqnarray}
where $(x,y,z)$ is a real vector called Bloch vector satisfying $%
x^{2}+y^{2}+z^{2}\leq 1.$ Any qubit state $\tau$ can be written in
the Bloch representation as
\begin{eqnarray}
\tau =\frac{1}{2}\left(
\begin{array}{cc}
1+z^{\prime } & x^{\prime }-iy^{\prime } \\
x^{\prime }+iy^{\prime } & 1-z^{\prime }%
\end{array}%
\right)   \label{III.2}
\end{eqnarray}
with $(x^{\prime },y^{\prime },z^{\prime })$ the Bloch vector satisfying $%
x^{\prime 2}+y^{\prime 2}+z^{\prime 2}\leq 1.$

Consequently, for $s\geq 0,$ we have
\begin{eqnarray}
&&\rho +s\tau  \nonumber \\
&=&\frac{1}{2}\left(
\begin{array}{cc}
1+s+z+sz^{\prime } & x+sx^{\prime }-i(y+sy^{\prime }) \\
x+sx^{\prime }+i(y+sy^{\prime }) & 1+s-(z+sz^{\prime })%
\end{array}%
\right) . \nonumber \\  \label{III.3}
\end{eqnarray}
$\rho +s\tau \geq 0$ implies
\begin{eqnarray}
y+sy^{\prime } &=&0,  \label{III.4} \\
x+sx^{\prime } &\geq &0.  \label{III.5}
\end{eqnarray}
Then we need to find the minimum $s\geq 0$ for given $(x,y)$ and possible $%
(x^{\prime },y^{\prime })$ under the conditions Eqs. (\ref{III.4},\ref{III.5}) and
\begin{eqnarray}
x^{2}+y^{2} &\leq &1,  \label{III.6} \\
x^{\prime 2}+y^{\prime 2} &\leq &1.  \label{III.7}
\end{eqnarray}

We consider different situations of $(x,y).$

(1). $x\geq 0,$ $y=0.$ For this case, $\rho \geq 0,$ then $s=0.$

(2). $x\geq 0,$ $y\neq 0.$ let $x^{\prime }\geq 0,$ then Eq. (\ref{III.5}) holds for
any $s\geq 0.$ Eq. (\ref{III.4}) implies $s=|\frac{y}{y^{\prime }}|.$ To minimize $s,$%
let $|y^{\prime }|=1,$ thus $s=|y|.$

(3). $x<0,y=0.$ let $y^{\prime }=0,$ then Eq. (\ref{III.4}) holds for any $s\geq 0.$
Eq. (\ref{III.5}) yields $x^{\prime }>0$ and
$s\geq |\frac{x}{x^{\prime }}|.$ To minimize $s,$ let $x^{\prime }=1,$ then $%
s=|x|.$

(4). $x<0,y\neq 0.$ Eq. (\ref{III.5}) yields $x^{\prime }>0$ and
\begin{eqnarray}
s\geq |\frac{x}{x^{\prime }}|. \label{III.8}
\end{eqnarray}
Eq. (\ref{III.4}) implies
\begin{eqnarray}
s=|\frac{y}{y^{\prime }}|. \label{III.9}
\end{eqnarray}
Eqs. (\ref{III.8},\ref{III.9}) imply
\begin{eqnarray}
s&=&|\frac{y}{y^{\prime }}|\geq |\frac{x}{x^{\prime }}|,  \label{III.10} \\
&&|\frac{y}{x}| \geq |\frac{y^{\prime }}{x^{\prime }}|,  \label{III.11}
\end{eqnarray}
To minimize $s,$ let $|x^{\prime }|=\frac{|x|}{\sqrt{x^{2}+y^{2}}},$ $|y^{\prime
}|=\frac{|y|}{\sqrt{x^{2}+y^{2}}},$ then $s=\sqrt{x^{2}+y^{2}}.$

In conclusion, we get
\begin{eqnarray}
\mathcal{P}_{\text{rob}}(\rho )=
   \begin{cases}
\ \ \ \ \ \ |y|, & x\geq 0, \\
\sqrt{x^{2}+y^{2}}, & x< 0.
   \end{cases}   \label{III.12}
\end{eqnarray}

\section{Proof of Theorem 4: $\mathcal{P}_{1}(\protect\rho )$ is a phase measure under
$C_{F}^{\prime }$}
\setcounter{equation}{0} \renewcommand\theequation{IV.\arabic{equation}}
For the complex number $z\in C,$ we write $z=x+iy=|z|e^{i\theta }$ with $x,y,\theta \in R,$
 $|z|=\sqrt{x^{2}+y^{2}}$. Define the function $\xi (z)$ as
\begin{eqnarray}
\xi (z)=|z|-x=|z|(1-\cos \theta )=2|z|\sin ^{2}\frac{\theta }{2}. \label{IV.1}
\end{eqnarray}
$\xi (z)$ has the following properties: for complex numbers $z,$ $z_{1}$ and
$z_{2},$
\begin{eqnarray}
\xi (z) &\geq &0\text{ and }\xi (z)=0\Leftrightarrow z\geq 0;  \label{IV.2} \\
\xi (tz) &=&t\xi (z),t\geq 0;  \label{IV.3} \\
\xi (z^{\ast }) &=&\xi (z);   \label{IV.4} \\
\xi (z_{1}z_{2}^{\ast }) &=&|z_{1}z_{2}|-z_{1}\cdot
z_{2}=|z_{1}z_{2}|(1-\cos \alpha ), \nonumber \\
&&\cos \alpha =\frac{z_{1}\cdot z_{2}}{%
|z_{1}z_{2}|};  \label{IV.5} \\
\xi (z_{1}z_{2}^{\ast }) &\geq &0,``="\Leftrightarrow z_{1}\cdot z_{2}=|z_{1}z_{2}|.   \label{IV.6} \\
\xi (z_{1}+z_{2}) &\leq &\xi (z_{1})+\xi (z_{2}),``="\Leftrightarrow
z_{1}\cdot z_{2}=|z_{1}z_{2}|.  \nonumber \\  \label{IV.7}
\end{eqnarray}
Write $z_{1}=x_{1}+iy_{1}$ and $z_{2}=x_{2}+iy_{2}$ with $%
\{x_{1},y_{1},x_{2},y_{2}\}\subset R.$ In Eqs. (\ref{IV.5},\ref{IV.6},\ref{IV.7}), $z_{1}\cdot z_{2}$ is the inner product $z_{1}\cdot z_{2}=x_{1}x_{2}+y_{1}y_{2}.$ Eq. (\ref{IV.7}) holds since
\begin{eqnarray}
&&\xi (z_{1}+z_{2}) \nonumber \\
&=&\xi ((x_{1}+x_{2})+i(y_{1}+y_{2}))  \nonumber \\
&=&\sqrt{(x_{1}+x_{2})^{2}+(y_{1}+y_{2})^{2}}-(x_{1}+x_{2})  \nonumber \\
&\leq &(\sqrt{x_{1}^{2}+y_{1}^{2}}-x_{1})+(\sqrt{x_{2}^{2}+y_{2}^{2}}-x_{2}), \label{IV.8}
\end{eqnarray}
and equality if and only if $z_{1}$, $z_{2}$ have the same direction, i.e., $z_{1}\cdot z_{2}=|z_{1}z_{2}|.$

Now we turn to $\mathcal{P}_{1}(\rho ).$ By definition,
\begin{eqnarray}
\mathcal{P}_{1}(\rho )=\sum_{j,k}\xi (\rho _{jk}), \label{IV.9}
\end{eqnarray}
then Eq. (\ref{IV.2}) implies (P1), Eq. (\ref{IV.7}) implies (P4). We only need to prove that $\mathcal{P}_{1}(\rho )$
satisfies (P3) under $C_{F}^{\prime }.$ Note that the definition of $\mathcal{P}_{1}(\rho
)$ can be directly extended to general positive semidefinite matrices, not
necessarily of unit trace, that is, $\mathcal{P}_{1}(t\rho )=t\mathcal{P}_{1}(\rho )$ for $t\geq 0.$

Suppose a channel $\phi \in C_{F}^{\prime }.$ Applying Theorem 1, $\phi $
can be expressed by the Kraus operators $\phi =\{Z_{\lambda }\}_{\lambda
}\cup \{M_{\mu }\}_{\mu }$ such that $\{Z_{\lambda }\}_{\lambda }\in O_{0}$
and $\{M_{\mu }\}_{\mu }\in O_{2}.$ For any state $\rho ,$ since $Z_{\lambda
}\rho Z_{\lambda }^{\dagger }\geq 0$ and $\mathcal{P}_{1}(\frac{Z_{\lambda }\rho
Z_{\lambda }^{\dagger }}{\text{tr}(Z_{\lambda }\rho Z_{\lambda }^{\dagger })}%
)=0,$ then we only need to prove that
\begin{eqnarray}
\sum_{\mu }\mathcal{P}_{1}(M_{\mu }\rho M_{\mu }^{t})\leq \mathcal{P}_{1}(\rho ). \label{IV.10}
\end{eqnarray}

Expanding $M_{\mu }$ and $\rho $ in the fixed orthonormal basis $\{|j\rangle \}_{j=1}^{d}$ as $M_{\mu }=\sum_{jk}M_{\mu
,jk}|j\rangle \langle k|$ and $\rho =\sum_{jk}\rho _{jk}|j\rangle \langle
k|, $ then
\begin{eqnarray}
&&M_{\mu }\rho M_{\mu }^{\dagger } =\sum_{jk,lm}M_{\mu ,jl}\rho _{lm}M_{\mu
,km}|j\rangle \langle k|,  \label{IV.11} \\
&&\sum_{\mu }\mathcal{P}_{1}(M_{\mu }\rho M_{\mu }^{\dagger }) \nonumber \\
&=&\sum_{\mu ,jk}\xi
(\sum_{lm}M_{\mu ,jl}\rho _{lm}M_{\mu ,km}) \nonumber \\
&\leq &\sum_{\mu ,jk}\sum_{lm}M_{\mu ,jl}M_{\mu ,km}\xi (\rho _{lm})  \label{IV.12}  \\
&=&\sum_{lm}[\sum_{\mu }(\sum_{j}M_{\mu ,jl})(\sum_{k}M_{\mu ,km})]\xi (\rho
_{lm}) \nonumber  \\
&\leq &\sum_{lm}\sqrt{[\sum_{\mu }(\sum_{j}M_{\mu ,jl})^{2}][\sum_{\mu
}(\sum_{k}M_{\mu ,km})^{2}]}\xi (\rho _{lm}) \nonumber \\  \label{IV.13} \\
&\leq &\sum_{lm}\xi (\rho _{lm})=\mathcal{P}_{1}(\rho ).  \label{IV.14}
\end{eqnarray}
In inequality (\ref{IV.12}) we have used Eqs. (\ref{IV.7},\ref{IV.3}). In inequality (\ref{IV.13}) we have used Cauchy-Schwarz
inequality. In inequality (\ref{IV.14}) we have used the fact
\begin{eqnarray}
\sum_{\mu }(\sum_{j}M_{\mu ,jl})^{2}\leq 1.  \label{IV.15}
\end{eqnarray}%
Since $\{M_{\mu }\}_{\mu }\in O_{2},$ then $\{M_{\mu ,jl}\}_{j=1}^{d}$ has
at most one nonzero element (which is positive), $(\sum_{j}M_{\mu
,jl})^{2}=\sum_{j}M_{\mu ,jl}^{2},$ and $\sum_{\mu }M_{\mu }^{\dagger
}M_{\mu }\preceq I$ results in inequality (\ref{IV.15}). Thus we proved (P3).

For qubit state $\rho $ expressed in Eq. (\ref{III.1}), $\mathcal{P}_{1}(\rho )$ reads
\begin{eqnarray}
\mathcal{P}_{1}(\rho )=\sqrt{x^{2}+y^{2}}-x.
\end{eqnarray}


\end{document}